# Quantum Mechanics helps in searching for a needle in a haystack

Lov K. Grover
*3C-404A Bell Labs, 600 Mountain Avenue, Murray Hill NJ 07974, lkgrover@bell-labs.com*

Quantum mechanics can speed up a range of search applications over unsorted data. For example imagine a phone directory containing $N$ names arranged in completely random order. To find someone's phone number with a probability of $50\%$, any classical algorithm (whether deterministic or probabilistic) will need to access the database a minimum of $0.5N$ times. Quantum mechanical systems can be in a superposition of states and simultaneously examine multiple names. By properly adjusting the phases of various operations, successful computations reinforce each other while others interfere randomly. As a result, the desired phone number can be obtained in only $O(\sqrt{N})$ accesses to the database.

**1. Introduction** In 1994 Shor discovered a quantum mechanical algorithm for factorization that was exponentially faster than any known classical algorithm [1]. This paper presents a quantum mechanical algorithm for search that is only polynomially faster than any classical algorithm; however, it does not depend for its impact on the unproven difficulty of the factorization problem. The search problem is this: there is an unsorted database containing $N$ items out of which just one item satisfies a given condition - that one item has to be retrieved. Once an item is examined, it is possible to tell whether or not it satisfies the condition in one step. However, there does not exist any sorting on the database that would aid its selection. The most efficient classical algorithm for this is to examine the items in the database one by one. If an item satisfies the required condition stop; if it does not, keep track of this item so that it is not examined again. It is easily seen that this algorithm will need to examine an average of $0.5N$ items before finding the desired item.

It is possible for quantum mechanical systems to make *interaction-free measurements* by using the duality properties of photons [2]. In these the presence (or absence) of an object can be deduced by allowing a small probability of a photon interacting with the object. Therefore, most probably the photon will not interact, however, just allowing a small probability of interaction is enough to make the measurement. Thus in the search problem also, it might be possible to find the object without examining all of the objects, but just by allowing a certain probability of examining the desired object.

Indeed, this paper shows that by using the same amount of hardware as in the classical case, but by having the input and output in *superpositions* of states, we can find an object in $O(\sqrt{N})$ *quantum mechanical steps* instead of $O(N)$ classical steps. Each *quantum mechanical step* consists of an elementary unitary operation (discussed in the following paragraph).

**1.1 Quantum Mechanical Algorithms** In a quantum computer, the logic circuitry and time steps are essentially classical, only the memory *bits* that hold the variables are in quantum superpositions (see [1] & [3] for a more detailed introduction to quantum computers). Quantum mechanical operations that can be carried out in a controlled way are unitary operations that act on a small number of bits in each step. The quantum search algorithm of this paper is a sequence of such unitary operations on a pure state, followed by a measurement operation. The three elementary unitary operations needed are the following. First is the creation of a superposition in which the amplitude of the system being in any of the $N$ basic states of the system is equal; second is the Walsh-Hadamard transformation operation and third the selective rotation of the phases of states.

A basic operation in quantum computing is the operation $M$ performed on a single bit that is represented by the following matrix: $M = \frac{1}{\sqrt{2}}\begin{bmatrix} 1 & 1 \\ 1 & -1 \end{bmatrix}$, i.e. a bit in the state 0 is transformed into a superposition in the two states: $\left(\frac{1}{\sqrt{2}}, \frac{1}{\sqrt{2}}\right)$. Similarly a bit in the state 1 is transformed into $\left(\frac{1}{\sqrt{2}}, -\frac{1}{\sqrt{2}}\right)$, i.e. the magnitude of the amplitude in each state is $\frac{1}{\sqrt{2}}$ but the *phase* of the amplitude in the state 1 is inverted. The phase does not have an analog in classical probabilistic algorithms. It comes about in quantum mechanics since the amplitudes are in general complex. In a system in which the states are described by $n$ bits (it has $N = 2^n$ possible states) we can perform the transformation $M$ on each bit independently in sequence thus changing the state of the system. The state transition matrix representing this operation will be of dimension $2^n$ X $2^n$. In case the initial configuration was the configuration with all $n$ bits in the first state, the resultant configuration will have an identical amplitude of $2^{-\frac{n}{2}}$ in each of the $2^n$ states. This is a way of creating a superposition with the same amplitude in all $2^n$ states.

Next consider the case when the starting state is another one of the $2^n$ states, i.e. a state described by an $n$ bit binary string with some 0s and some 1s. The result of performing the transformation $M$ on each bit will be a superposition of states described by all possible $n$ bit binary strings with amplitude of each state having a magnitude equal to $2^{-\frac{n}{2}}$ and sign either + or -. To deduce the sign, observe that from the definition of the matrix $M$, i.e. $M = \frac{1}{\sqrt{2}}\begin{bmatrix} 1 & 1 \\ 1 & -1 \end{bmatrix}$, the phase of the resulting configuration is changed when a bit that was previously a 1 remains a 1 after the transformation is performed. Hence if $\bar{x}$ be the $n$-bit binary string describing the starting state and $\bar{y}$ the $n$-bit binary string describing the resulting string, the sign of the amplitude of $\bar{y}$ is determined by the parity of the bitwise dot product of $\bar{x}$ and $\bar{y}$, i.e. $(-1)^{\bar{x} \cdot \bar{y}}$. This transformation is referred to as the Walsh-Hadamard transformation [4]. This operation (or a closely related operation called the Fourier Transformation) is one of the things that makes quantum mechanical algorithms more powerful than classical algorithms and forms the basis for most significant quantum mechanical algorithms.

The third transformation that we will need is the selective rotation of the phase of the amplitude in certain states. The transformation describing this for a 2 state system is of the form: $\begin{bmatrix} e^{j\phi_1} & 0 \\ 0 & e^{j\phi_2} \end{bmatrix}$, where $j = \sqrt{-1}$ and $\phi_1, \phi_2$ are arbitrary real numbers. Note that, unlike the Walsh-Hadamard transformation and other state transition matrices, the probability in each state stays the same since the square of the absolute value of the amplitude in each state stays the same.

## 2. The Abstracted Problem

Let a system have $N = 2^n$ states which are labelled $S_1, S_2, ... S_N$. These $2^n$ states are represented as $n$ bit strings. Let there be a unique state, say $S_v$, that satisfies the condition $C(S_v) = 1$, whereas for all other states $S$, $C(S) = 0$ (assume that for any state $S$, the condition $C(S)$ can be evaluated in unit time). The problem is to identify the state $S_v$.

This could represent a database search problem where the function $C(S)$ is based on the contents of memory location corresponding to state $S$ (as discussed in the abstract). Alternatively it could represent a problem where the function $C(S)$ was being evaluated by the computer. Various important computer science problems can be represented in this form [3] [5] [9].

## 3. Algorithm

Steps (i) & (ii) are a sequence of elementary unitary operations of the type discussed in section 1.1. Step (iii) is the final *measurement* by an external system.

(i) Initialize the system to the superposition:
$\left(\frac{1}{\sqrt{N}}, \frac{1}{\sqrt{N}}, \frac{1}{\sqrt{N}} ... \frac{1}{\sqrt{N}}\right)$, i.e. there is the same amplitude to be in each of the $N$ states. This superposition can be obtained in $O(\log N)$ steps, as discussed in section 1.1.

(ii) Repeat the following unitary operations $O(\sqrt{N})$ times (the precise number of repetitions is important as discussed in [5]):
   (a) Let the system be in any state S:
      In case $C(S) = 1$, rotate the phase by $\pi$ radians;
      In case $C(S) = 0$, leave the system unaltered.
   (b) Apply the diffusion transform $D$ which is defined by the matrix $D$ as follows:
$D_{ij} = \frac{2}{N}$ if $i \neq j$ & $D_{ii} = -1 + \frac{2}{N}$.
($D$ can be implemented as a product of 3 elementary matrices as discussed in section 5).

(iii) Measure the resulting state. This will be the state $S_v$ (i.e. the desired state that satisfies the condition $C(S_v) = 1$) with a probability of at least 0.5.

Note that step (ii) (a) is a phase rotation transformation of the type discussed in the last paragraph of section 1.1. In an implementation it would involve a portion of the quantum system sensing the state and then deciding whether or not to rotate the phase. It would do it in a way so that no trace of the state of the system be left after this operation so as to ensure that paths leading to the same final state were indistinguishable and could interfere. Reference [5] gives a way of doing this with a single quantum query. Note that this does *not* involve a classical measurement.

**4. Convergence** The loop in step (ii) above, is the heart of the algorithm. Each iteration of this loop increases the amplitude in the desired state by $O\left(\frac{1}{\sqrt{N}}\right)$, as a result in $O(\sqrt{N})$ repetitions of the loop, the amplitude and hence the probability in the desired state reach $O(1)$. In order to see that the amplitude increases by $O\left(\frac{1}{\sqrt{N}}\right)$ in each repetition, we first show that the diffusion transform, $D$, can be interpreted as an *inversion about average* operation. Just a simple inversion is a phase rotation operation and by the discussion in the last paragraph of section 1.1, is unitary. In the following discussion we show that the *inversion about average* operation (defined more precisely below) is also a unitary operation and is equivalent to the diffusion transform $D$ as used in step (ii)(a) of the algorithm.

Let $\alpha$ denote the average amplitude over all states, i.e. if $\alpha_i$ be the amplitude in the $i^{th}$ state, then the average is $\frac{1}{N}\sum_{i=1}^{N}\alpha_i$. As a result of the operation $D$, the amplitude in each state increases (decreases) so that after this operation it is as much below (above) $\alpha$, as it was above (below) $\alpha$ before the operation.

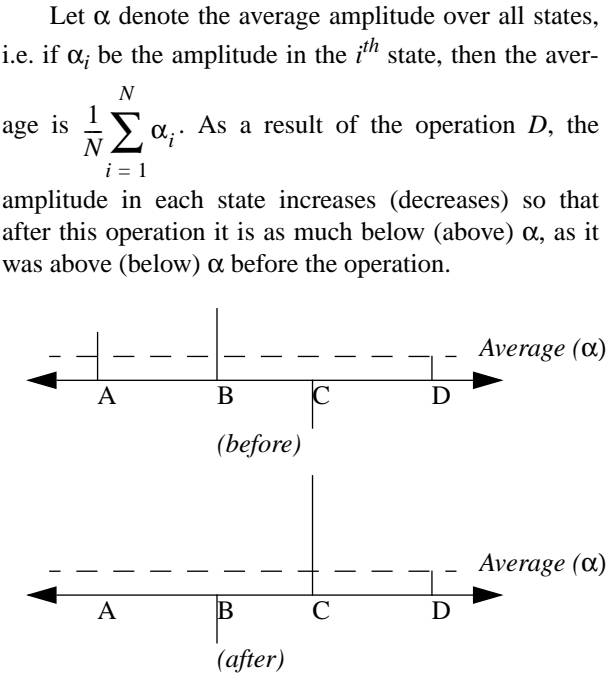

**Figure 1.** *Inversion about average* **operation.**

The diffusion transform, $D$, is defined as follows:

(4.0) $D_{ij} = \frac{2}{N}$, if $i \neq j$ & $D_{ii} = -1 + \frac{2}{N}$.

Observe that $D$ can be represented in the form $D \equiv -I + 2P$ where $I$ is the identity matrix and $P$ is a projection matrix with $P_{ij} = \frac{1}{N}$ for all $i, j$. The following two properties of $P$ are easily verified: first, that $P^2 = P$ & second, that $P$ acting on any vector $\bar{v}$ gives a vector each of whose components is equal to the average of all components.

Using the fact that $P^2 = P$, it follows immediately from the representation $D = -I + 2P$ that $D^2 = I$ and hence $D$ is unitary.

In order to see that $D$ is the *inversion about average*, consider what happens when $D$ acts on an arbitrary vector $\bar{v}$. Expressing $D$ as $-I + 2P$, it follows that: $D\bar{v} = (-I + 2P)\bar{v} = -\bar{v} + 2P\bar{v}$. By the discussion above, each component of the vector $P\bar{v}$ is $A$ where $A$ is the average of all components of the vector $\bar{v}$. Therefore the $i^{th}$ component of the vector $D\bar{v}$ is given by $(-v_i + 2A)$ which can be written as $(A + (A - v_i))$ which is precisely the *inversion about average*.

Next consider the situation in figure 2, when this operation is applied to a vector with each of the components, except one, having an amplitude equal to $\frac{C}{\sqrt{N}}$ where $C$ lies between $\frac{1}{2}$ & 1; the one component that is different has an amplitude of $-\sqrt{1-C^2}$.

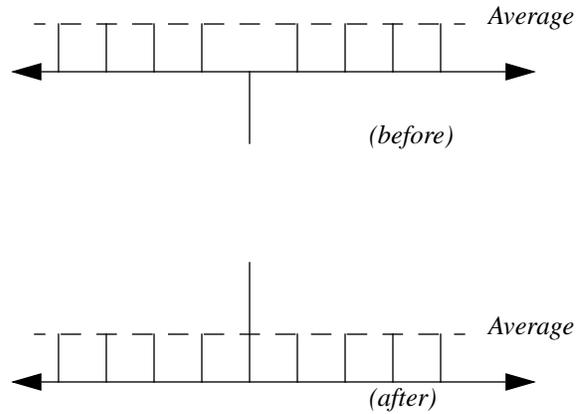

**Figure 2.** **I***nversion about average* **operation is applied to a superposition where all but one of the components are initially identical and of magnitude** $O\left(\frac{1}{\sqrt{N}}\right)$**.**

The average $A$ of all components is approximately equal to $\frac{C}{\sqrt{N}}$. Since each of the $(N-1)$ components is approximately equal to the average, they do not change significantly as a result of the inversion about average. The one component that was negative, now becomes positive and its magnitude increases by $\frac{2C}{\sqrt{N}}$.

In the loop of step (ii) of section 3, first the amplitude in a selected state is inverted (this is a phase rotation and hence a valid quantum mechanical operation as discussed in the last paragraph of section 1.1). Then the *inversion about average* operation is carried out. This increases the amplitude in the selected state in each iteration by $\frac{2C}{\sqrt{N}}$. Therefore as long as the magnitude of the amplitude in the single state, i. e. $\sqrt{1-C^2}$, is less than $\frac{1}{\sqrt{2}}$, the increase in its magnitude is greater than $\frac{1}{\sqrt{2N}}$. It immediately follows that there exists a number $M$ less than $\sqrt{N}$, such that in $M$ repetitions of the loop in step (ii), the magnitude of the amplitude in the desired state will exceed $\frac{1}{\sqrt{2}}$. Therefore if the state of the system is now measured, it will be in the desired state with a probability greater than 0.5.

**5. Implementation** As mentioned in section 1.1, quantum mechanical operations which can be implemented in terms of elementary unitary operations, are local transition matrices, i.e. matrices in which only a constant number of elements in each column are non-zero. The diffusion transform $D$ is defined in step (ii)(b) of the algorithm as: $D_{ij} = \frac{2}{N}$ if $i \neq j$ & $D_{ii} = -1 + \frac{2}{N}$.

$D$ as presented above, is not a local transition matrix since there are transitions from each state to all $N$ states. Using the Walsh-Hadamard transformation matrix (section 1.1), $D$ can be implemented as a product of three local unitary transformations as $D = WRW$, where $R$ the phase rotation matrix & $W$ the Walsh-Hadamard Transform Matrix are defined as follows: $R_{ij} = 0$ if $i \neq j$; $R_{ii} = 1$ if $i = 0$; $R_{ii} = -1$ if $i \neq 0$. $W_{ij} = 2^{-n/2}(-1)^{\bar{i}\cdot\bar{j}}$, $\bar{i}$ is the binary representation of $i$, and $\bar{i}\cdot\bar{j}$ denotes the bitwise dot product of the two $n$ bit strings $\bar{i}$ and $\bar{j}$.

Each of $W$ & $R$ is a local transition matrix. $R$ as defined above is a phase rotation matrix and is clearly local. $W$, when implemented as in section 1.1, is a local transition matrix on each bit.

We evaluate $WRW$ and show that it is indeed equal to $D$. $R$ can be written as $R = R_1 + R_2$ where $R_1 = -I$, $I$ is the identity matrix and $R_{2,00} = 2$, $R_{2,ij} = 0$ if $i \neq 0, j \neq 0$. By observing that $MM = I$ where $M$ is the matrix defined in section 1.1, it is easily proved that $WW=I$ and hence $D_1 = WR_1W = -I$. We next evaluate $D_2 = WR_2W$. By standard matrix multiplication: $D_{2,ad} = \sum_{bc} W_{ab}R_{2,bc}W_{cd}$. Using the definition of $R_2$ and the fact $N = 2^n$, it follows that $D_{2,ad} = 2W_{a0}W_{0d} = \frac{2}{2^n}(-1)^{\bar{a}\cdot\bar{0}+\bar{0}\cdot\bar{d}} = \frac{2}{N}$. Thus all elements of the matrix $D_2$ equal $\frac{2}{N}$, the sum of the two matrices $D_1$ and $D_2$ gives $D$.

The quantum search algorithm of this paper is likely to be simpler to implement as compared to many other known quantum mechanical algorithms. This is because the only operations required are the Walsh-Hadamard transform & the conditional phase shift operation, both of which are relatively easy as compared to operations required for other quantum mechanical algorithms [6]. Also, quantum mechanical algorithms based on the Walsh-Hadamard transform (e.g. the search algorithm of this paper, [4], [7], [8]) are likely to be much simpler to implement than those based on the "large scale Fourier transform" [1], [6].

**6. Acknowledgments** Peter Shor, Ethan Bernstein, Gilles Brassard, Norm Margolus & John Preskill for helpful comments.